\documentclass[aps,pra,twocolumn,superscriptaddress]{revtex4}
\usepackage{graphicx}
\usepackage{amssymb}
\usepackage{amsmath}
\usepackage{mathtools}
\usepackage{bm}

\begin{document}

\title{Emergence and scaling of spin turbulence in quenched antiferromagnetic spinor Bose-Einstein condensates}

\author{Seji Kang}
\affiliation{Department of Physics and Astronomy, and Institute of Applied Physics, Seoul National University, Seoul 08826, Korea}
\affiliation{Center for Correlated Electron Systems, Institute for Basic Science, Seoul 08826, Korea}

\author{Sang Won Seo}
\affiliation{Department of Physics and Astronomy, and Institute of Applied Physics, Seoul National University, Seoul 08826, Korea}
\affiliation{Center for Correlated Electron Systems, Institute for Basic Science, Seoul 08826, Korea}

\author{Joon Hyun Kim}
\affiliation{Department of Physics and Astronomy, and Institute of Applied Physics, Seoul National University, Seoul 08826, Korea}

\author{Y. Shin}\email{yishin@snu.ac.kr}
\affiliation{Department of Physics and Astronomy, and Institute of Applied Physics, Seoul National University, Seoul 08826, Korea}
\affiliation{Center for Correlated Electron Systems, Institute for Basic Science, Seoul 08826, Korea}


\begin{abstract}
We investigate the phase transition dynamics of a quasi-2D antiferromagnetic spin-1 Bose-Einstein condensate from the easy-axis polar phase to the easy-plane polar phase, which is initiated by suddenly changing the sign of the quadratic Zeeman energy $q$. We observe the emergence and decay of spin turbulence and the formation of half-quantum vortices (HQVs) in the quenched condensate. The characteristic time and length scales of the turbulence generation dynamics are proportional to $|q|^{-1/2}$ as inherited from the dynamic instability of the initial state. In the evolution of the spin turbulence, spin wave excitations develop from large to small length scales, suggesting a direct energy cascade, and the spin population for the axial polar domains exhibit  a nonexponential decay. The final equilibrated condensate contains HQVs, and the number is found to increase and saturate with increasing $|q|$. Our results demonstrate the time-space scaling properties of the phase transition dynamics near the critical point and the peculiarities of the spin turbulence state of the antiferromagnetic spinor condensate. 
\end{abstract}

\maketitle

\section{Introduction}

The far-from-equilibrium dynamics of many-body quantum systems is a challenging subject to study in modern physics~\cite{Review} and relevant to several areas from cosmology~\cite{Traschen90} to condensed matter physics~\cite{Fausti11}. Ultracold atomic gases provide a highly controllable platform for studying many-body physics~\cite{Bloch_rmp08} and a quench protocol is typically employed to explore nonequilibrium dynamics~\cite{Polkovnikov_rmp}, where a system is prepared in a well-defined initial state and then its evolution is precisely examined after the system's Hamiltonian is rapidly changed. One of the key topics in current research activities is quantum phase transition dynamics, which addresses the important question of how a many-body system evolves into a newly ordered quantum state. Recently, it was theoretically proposed that scaling behavior occurs in the phase transition dynamics near quantum critical points~\cite{Lamacraft_prl, Uhlmann_prl07, Damski_prl07, Rossini09, DallaTorre13, Karl_scirep} and the existence of scaling and universality was indeed demonstrated in various experiments of quantum phase transitions~\cite{Braun_pnas,Nicklas_prl,Anquez16,Clark_sci}.

In this paper, we report on an experimental study of a quantum phase transition of a quasi-2D antiferromagnetic spin-1 Bose-Einstein condensate (BEC). The ground state of the antiferromagnetic BEC is a polar state with $\langle \textbf{F} \rangle =0$, where $\textbf{F}=(F_x,F_y,F_z)$ is the hyperfine spin operator of the atoms~\cite{Kawaguchi_rev, StamperK_rev}. The spin state is parametrized with a unit vector $\vec{d}=(d_x,d_y,d_z)$, called a spin nematic director, such that the system is in the $|m_F=0\rangle$ state for the quantization axis along $\vec{d}$. In an external magnetic field $\vec{B}$ (e.g., along the $z$ direction), an uniaxial spin anisotropy is imposed on the system due to the quadratic Zeeman energy $E_Z=q\langle F_z^2\rangle=q(1-d_z^2)$ and the BEC has two ground-state phases depending on the sign of $q$: the easy-axis polar (EAP) phase with $\vec{d}\parallel \hat{z}$ for $q>0$ and the easy-plane polar (EPP) phase with $\vec{d}\perp \hat{z}$ for $q<0$. Here, we investigate the EAP-to-EPP phase transition dynamics by preparing a quasi-2D BEC with a uniform spin texture with $\vec{d}\parallel \hat{z}$ for positive $q$ [Fig.~1(a)] and then suddenly changing the sign of $q$ to negative. In the transition to the EPP phase, the continuous spin-rotation symmetry in the $xy$ plane is spontaneously broken and topological point defects, which are half-quantum vortices (HQVs) in the EPP phase~\cite{Zhou_prl01,Zhou_IJMPB03,Seo_prl15,Seo_prl16}, can be created in the spatially extended 2D system [Fig.~1(b)].

The dynamic instability of the initial EAP state for $q<0$ was demonstrated in experiments with elongated BECs~\cite{Bookjans_prl,Vinit_arxiv}. 
The Bogoliubov analysis of the EAP state gives two degenerate magnon modes with energy spectra of $E_k=\sqrt{(\epsilon_k +q)(\epsilon_k +q +2 c_2 n)}$~\cite{Kawaguchi_rev}, where $\epsilon_k=\hbar^2 k^2 /(2m)$ is the single-particle spectrum ($m$ is the atomic mass), $c_2$ is the spin interaction coefficient, and $n$ is the atomic density. For small negative $q$ such that $|q|\ll c_2n$, the magnon modes with $k <k_q\equiv\sqrt{2|q|m}/\hbar$ have imaginary frequencies for $E_k^2 <0$, which means that small fluctuations in the transverse magnetization would be exponentially amplified in the EAP state. The dynamic instability rate is given by the maximum magnitude of the imaginary frequencies, $\Gamma_q=2|E_{k=0}|/\hbar \approx \sqrt{8|q|c_2 n}/\hbar$. Thus, we can define the time and length scales of $t_q = 2\pi/\Gamma_q\propto |q|^{-1/2}$ and $l_q= 1/k_q\propto |q|^{-1/2}$ to characterize the instability of the initial EAP state after the quench. The primary focus of this work is to determine how the intrinsic time-space scaling of the initial state near the critical point at $q=0$ is inherited and transformed in the subsequent phase transition dynamics.

\begin{figure*}
\centering
\includegraphics[width=17.0cm]{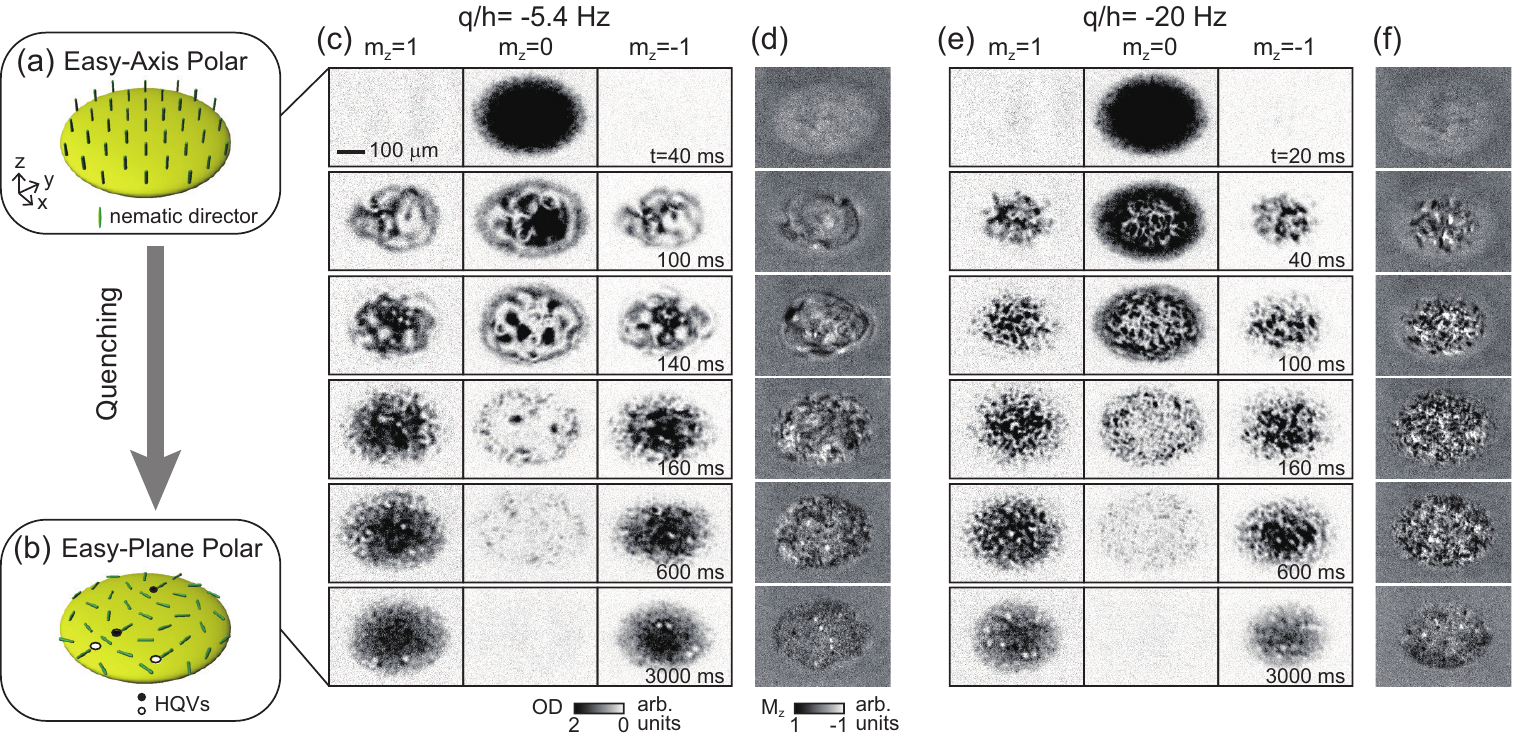}
\caption{Phase transition from the easy-axis polar (EAP) phase to the easy-plane polar (EPP) phase of an antiferromagnetic spin-1 Bose-Einstein condensate (BEC). (a) The spin nematic director $\vec{d}$ is initially aligned to the axial $z$ direction and (b) the phase transition to the EPP phase is triggered by a sudden change of the quadratic Zeeman energy, $q$. (c),(e) Absorption images of the $m_z=1,0,-1$ spin components after Stern-Gerlach (SG) spin separation and (d),(f) {\textit{in-situ}} axial magnetization images, $M_z(x,y)$, for various hold times $t$ after the quench to (c),(d) $q/h=-5.4$~Hz and (e),(f) $-20$~Hz. Half-quantum vortices (HQVs) are identified with density-depleted holes in the SG images and their magnetized cores in the $M_z$ image~\cite{Seo_prl15,Seo_prl16}.}
\label{fig1}
\end{figure*}

We investigate the scaling properties of the phase transition dynamics by measuring the spin texture evolution of the quenched BEC. We observe that spin turbulence emerges, which is featured by an irregular spin texture involving both a spatially disordered pattern of $\vec{d}$ and ferromagnetic spin excitations. We show that the turbulence generation dynamics are characterized by the time-space scaling of $\sim |q|^{-1/2}$ predicted for the initial EAP state. Furthermore, in the evolution of the spin turbulence, we observe that its spatial structure develops from large to small length scales, suggesting a direct energy cascade of spin wave excitations, and that the spin population for the axial polar domains exhibits a nonexponential decay. Finally, we observe that HQVs are created in the phase transition dynamics and find that the HQV number of the final equilibriated sample increases and saturates as $|q|$ increases.

The paper is organized as follows. In Sec.~II, we describe the experimental procedures for sample preparation and spin texture imaging. In Sec.~III, we present the characterization of the emergence and decay of spin turbulence in the quenched BEC and discuss the scaling properties of the phase transition dynamics. A summary and outlook are provided in Sec.~IV.

\section{Experiments}

Our experiment begins by preparing a BEC of $^{23}$Na atoms in the $|F=1, m_F=0\rangle$ state~\cite{Seo_prl15}. The condensate is confined in an optical potential with trapping frequencies of $(\omega_x, \omega_y, \omega_z)=2\pi\times(3.8, 5.5, 400)$~Hz. The condensate contains $N^c\approx 8.0\times 10^6$ atoms and its Thomas-Fermi radii are $(R_x, R_y, R_z)\approx(232, 160, 2.2)~\mu$m. The sample preparation is carried out in a magnetic field of $B_z=0.5$~G. Initially, the sample only has the $m_z=0$ spin component and its thermal fraction is less than 10\%. To initiate the transition dynamics to the EPP phase, we first adiabatically ramp $B_z$ to 33~mG for 0.2~s and then, we suddenly turn on a microwave field to change the quadratic Zeeman energy to a negative $q$ value~\cite{Gerbier_pra, Zhao_pra}.  The residual field gradient was measured to be less than 0.1~mG/cm.

In our experiment, $q/h$ ranges from $-1.4$~Hz to $-20$~Hz. The excitation energy of the initial state after the quench is given by $|q|$ with respect to the EPP ground state, which is much smaller than the condensate chemical potential $\mu= h\times880$~Hz. Therefore, the subsequent evolution of the condensate can be approximated as pure spin dynamics, and does not involve density excitations. For the peak atomic density, $n$, of the sample, the spin interaction energy is $c_2 n/h= 14$~Hz~\cite{Black_prl07} and the spin healing length is $\xi_s=\hbar/\sqrt{2m c_2 n}\approx 4.0~\mu$m. Because the condensate thickness $R_z$ is smaller than $\xi_s$, the spin dynamics in the oblate condensate are effectively 2D.

The spin texture of the condensate is examined by taking an absorption image after the Stern-Gerlach (SG) spin separation. After releasing the trapping potential, we apply a magnetic field gradient pulse along the $x$ direction and let the three $m_z=1,0,-1$ spin components be spatially separated for a 24-ms time of flight. The image reveals the density distributions of the individual spin components. The applied field gradient was slightly inhomogeneous and the expansions of the $m_z=\pm1$ spin components are not perfectly identical. We used $F=1$ imaging and calibrated the absorption coefficients for each spin component~\cite{Kim16}.  

Spin-sensitive {\it{in-situ}} phase-contrast imaging is also employed to obtain further information on the spatial magnetization structure of the condensate~\cite{Seo_prl15}. The probe beam frequency is tuned to give a signal proportional to the axial magnetization, $M_z$, i.e., the density difference of the $m_z=\pm 1$ spin components. The contribution of the $m_z=0$ spin component to the imaging signal is not significant and we interpret the phase-contrast image as the axial magnetization distribution $M_z(x,y)$ of the condensate.

\section{Results}

\subsection{Emergence of spin turbulence}

Figure~1 displays two image data sequences of the quenched BEC for the two different $q$ values of $-5.4$~Hz and $-20$~Hz. After a short hold time, an irregular spin texture begins to appear in the condensate. It is clearly shown that the $m_z=\pm1$ components are spatially separated from the $m_z=0$ component [Figs.~1(c) and 1(e)], which results from the immiscibility of the $m_z=\pm1$ components with the $m_z=0$ component~\cite{Stenger_nat}. The spin domains formed by an equal mixture of the $m_z=\pm1$ components have $\vec{d}\perp \hat{z}$, where the azimuthal direction of $\vec{d}$ is determined by the relative phase of the two spin components. It is observed that the irregular spin texture first emerges in the center region of the condensate and expands over the whole condensate. We attribute this result to the inhomogeneous density distribution of the trapped condensate because $\Gamma_q \propto \sqrt{n}$. 

The appearance of an irregular spin texture is also observed in the magnetization image [Figs.~1(d) and 1(f)]. In the spin-exchange process where two $m_z=0$ atoms are scattered into a pair of $m_z=+1$ and $-1$ atoms, the quadratic Zeeman energy is converted into the kinetic energy of the $m_z=\pm1$ atoms, imparting opposite momenta to the pair. Thus, spin currents are generated in the $m_z=\pm 1$ spin domains and axial magnetization develops at the domain boundaries. The irregular structure in the $M_z$ image constitutes an observation of spin turbulence that has a complex spin current pattern.

As the hold time $t$ increases, the $m_z=0$ spin component is continuously depleted, and the spin texture becomes more complex. In particualr, the length scale of the spin texture decreases, implying a direct energy cascade in the spin turbulence. The condensate eventually relaxes into the EPP phase with the $m_z=0$ component vanising. In the final state, HQVs are observed as magnetized point defects in the $M_z$ image~\cite{Seo_prl15} and are identified with the density-depleted holes in the SG image.

\begin{figure}
\centering
\includegraphics[width=6.5cm]{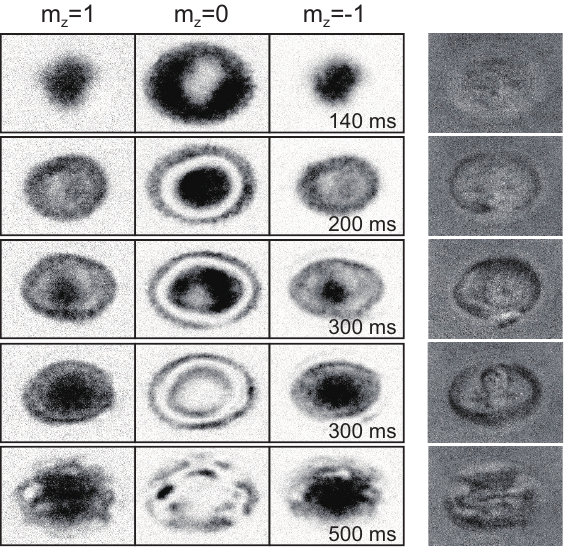}
\caption{Spin wave excitations for $q/h=-1.4$~Hz. SG images (left) and {\textit{in-situ}} magnetization images (right) for various hold times. A ring-shaped oscillating spin texture appears and shortly becomes dismantled.}
\label{fig2}
\end{figure}

It is apparent in the comparison between the two image data sets in Fig.~1 for $q/h=-5.4$~Hz and $-20$~Hz that when the system is closer to the critical point, the time and length scales of the quench dynamics become slower and larger, respectively, which is consistent with the theoretical anticipation based on the dynamic instability of the initial state. The length scale of the spin texture becomes even larger with lower $|q|$. For $|q/h|< 2$~Hz, we observed that the incipient spin texture shows a large ring-shaped pattern, which propagates toward the boundary and shortly becomes dismantled (Fig.~2). The ring-shaped pattern has the same ellipticiy as the trapped condensate and we believe that it corresponds to long-wavelength spin wave excitations induced by the trapping geometry of the finite-size sample~\cite{Klempt_prl,Scherer_prl}.

\subsection{Characterization of the quench dynamics}

\subsubsection{Time evolutions of $\eta$ and $\delta M_z^2$}

We first characterize the quench dynamics of the condensate by measuring the time evolutions of the fractional population, $\eta$, of the $m_z=0$ component and the magnetization variance, $\langle \delta M_z^2 \rangle$, of the spin texture (Fig.~3). Here $\eta= N^c_0/N^c$ and $ N^c=\sum_i N^c_i$, where $N^c_i$ is the $m_z=i$ atom number of the condensate ($i=1,0$, and $-1$) and is determined from the SG absorption image. The thermal cloud contribution is subtracted using a Gaussian fit to the outer thermal wing. For the measurement of $\langle \delta M_z^2 \rangle$, we set the central $206~\mu \textrm{m}\times206~\mu\textrm{m}$ region of the condensate as the region of interest.

\begin{figure}
	\centering
	\includegraphics[width=7.0cm]{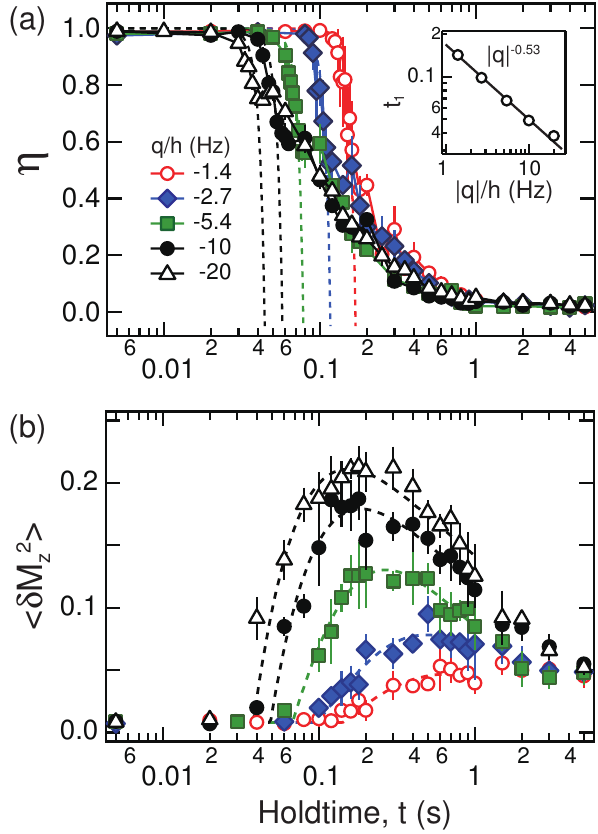}
	\caption{Temporal evolutions of (a) the fractional population, $\eta$, of the $m_z=0$ spin component in the condensate and (b) the magnetization variance, $\langle\delta M_z^2\rangle$, for various $q$ values. Each data point was obtained by averaging about five measurements and its error bar denotes the standard deviation of the measurements. The dashed lines are guide lines for eyes. The inset in (a) shows the time $t_1$ for $\eta(t_1)=0.8$ as a function of $|q|$. The solid line denotes a power-law fit to the data.}
\end{figure}

The quench evolution starts with a delayed, rapid decrease of $\eta$ as expected from the exponential growth of the dynamically unstable magnon modes. The early evolution of $\eta$ is described as $\eta(t)=1-b \exp(-\Gamma_q t)$, where $b$ is a constant determined by the magnitude of the magnetization fluctuations in the system. In the inset of Fig.~2(a), we display the time $t_1$ measured for $\eta(t_1)=0.8$ as a function of $|q|$. A power-law fit to the experimental data gives an exponent of $-0.53\pm 0.01$, which is in quantitatively good agreement with the predicted scaling of $\Gamma_q \sim |q|^{1/2}$. The measured $t_1$ values give $b\approx 4.6 \times 10^{-6}$, which is slightly higher than the value of $b\approx 3.3\times 10^{-6}$ estimated for quantum fluctuations at our peak atomic density~\cite{Mele_pra13}, indicating thermal enhancement in the experiment.

The rapid decay of $\eta$ is halted at a certain threshold value $\eta_{th}$ and after a short loitering period of a few tens of ms, $\eta$ resumes its decay. As $\langle \delta M_z^2\rangle$ rapidly increases when $\eta$ undergoes this change, it is reasonable to infer that the condensate enters a qualitatively different phase of the quench evolution, where the role of the generated spin turbulence becomes significant. The threshold value $\eta_{th}$ monotonically increases from $\approx 0.4$ for $q/h=-1.4$~Hz to $\approx 0.75$ for $q/h=-20$~Hz [Fig.~6(c)].

The spin turbulence develops further with maximizing $\langle \delta M_z^2 \rangle$ and then gradually relaxes with decreasing $\langle \delta M_z^2 \rangle$. After a long hold time, $t>5$~s, the system is equilibrated with $\eta\simeq 0$ and stationary $\langle \delta M_z^2 \rangle$. In our experiment, the equilibrium value of $\langle \delta M_z^2 \rangle$ was insensitive to $q$ because the final sample temperature is mainly determined by the heating from the microwave field dressing and the evaporation cooling due to the finite trap depth. Note that for $q/h=-1.4$~Hz, $\langle \delta M_z^2 \rangle$ monotonically increases and saturates to the equilibrium value over time. At the equilibrium, the thermal fraction was approximately 30\% and the sample temperature was estimated to be $T\approx 100$~nK. $k_B T \ll |q|$ and the thermal cloud was an equal mixture of the three spin components~\cite{Erhard_pra}.

\subsubsection{Power spectrum of axial magnetization}

To investigate the spatial structure of the generated spin turbulence, we measure the power spectrum of the magnetization distribution, $P(\vec{k})=|\int dr^2 e^{i\vec{k}\cdot\vec{r}}M_z(\vec{r})|^2$. In the measurement, a reference image was first obtained by averaging over ten images that were taken for the same experiment and we subtracted the reference image from the individual images to remove systematic fringes not related with the spin texture of the sample. Then, the power spectrum $P(\vec{k})$ was obtained by averaging the squared Fourier transforms of the subtracted images and subtracting the photon shot noise level. We introduce a relative spectrum $\tilde{P}(\vec{k})=P(\vec{k})/P_\textrm{eq}(\vec{k})$, where $P_\textrm{eq}(\vec{k})$ is the average spectrum of samples at thermal equilibrium in the EPP phase. Because $P_\textrm{eq}$ is measured with the same imaging system, the relative spectrum $\tilde{P}$ provides spectral information free from the systematic modifications of the imaging system. For the determination of the equilibrium spectrum $P_\textrm{eq}$, we selected samples for $q/h=-1.4$~Hz at $t=5$~s, in particular, without HQVs because HQVs can affect the spectrum due to their magnetized cores and the magnon excitations generated by their collisional motions~\cite{Seo_prl16}. $\tilde{P}(\vec{k})$ was isotropic and we obtained a 1D spectrum, $\tilde{P}(k)$, by azimuthally averaging it over $|\vec{k}|=k$.

\begin{figure}
\centering
\includegraphics[width=8.4cm]{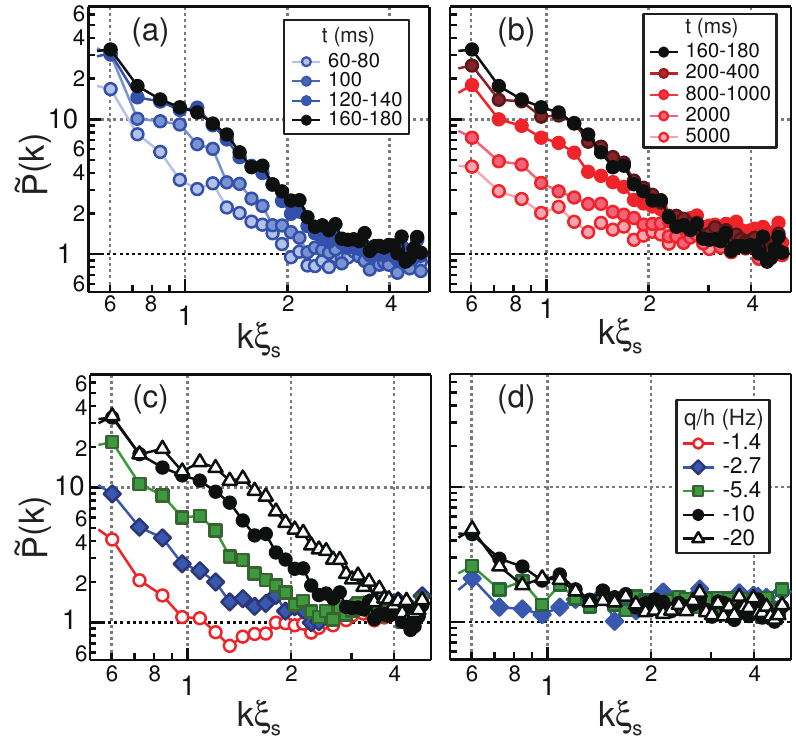}
\caption{Relative power spectra, $\tilde{P}(k)$, of the axial magnetizaion, $M_z(x,y)$, of the quenched BEC. Evolution of $\tilde{P}(k)$ for $q/h=-10$~Hz when $\langle \delta M_z^2\rangle$ increases (a) and decreases (b) [Fig.~3(b)]. (c) $\tilde{P}(k)$ at the time when $\langle \delta M_z^2\rangle$ is maximum and (d) at $t=5$~s when the quenched BEC is thermally equilibrated. The reference spectrum $\tilde{P}(k)=1$ was obtained from the thermal equilibrium samples at $T\approx 100$~nK, having no HQVs (see text).}
\label{fig2}
\end{figure}

Figures~4(a) and 4(b) show the evolution of $\tilde{P}(k)$ for $q/h=-10$~Hz when $\langle \delta M_z^2\rangle$ increases and decreases, respectively. As observed in the visual examination of Figs.~1(d) and 1(f), the $\tilde{P}$ measurement results show that the spin turbulence develops from low to high wave numbers $k$, i.e., from large to small length scales [Fig.~4(a)].  When $\langle \delta M_z^2\rangle$ decreases, the power spectrum decays towards the equilibrium level, $\tilde{P}_\textrm{eq}=1$, where the spectral strength subsides more rapidly in the lower-$k$ region [Fig.~4(b)]. Over the entire growth and decay evolution, the spectral center-of-mass of $\tilde{P}(k)$ continues moving towards high $k$, which suggests a direct energy cascade of the spin wave excitations in the spin turbulence. 

In Fig.~4(a), it is noted that when the spin turbulence is generated, the spectral slope is formed in $\tilde{P}(k)$ and propagates to the high-$k$ regions in a self-similar manner. In Fig.~4(c), we display $\tilde{P}(k)$ at the time when $\langle\delta M_z^2\rangle$ is maximum for various $q$ values and observe that the spectral slopes are almost identical, whereas the characteristic wave number increases with increasing $|q|$. The power-law scaling in the spin turbulence of antiferromagnetic BECs was predicted by Fujimoto~\textit{et al.}~\cite{Fujimoto_pra12, Fujimoto_pra13}, but further analysis of our experimental data is limited by a lack of quantitative understanding of $P_\textrm{eq}(k)$. In Ref.~\cite{Symes_pra}, Symes {\textit{et al.}} calculated the static structure factor of the antiferromagnetic BEC in the EPP phase at low temperatures. However, our sample temperature of $k_B T/(c_2 n)\approx 150$ is too high to be extrapolated from their results.

In Fig.~4(d), we display $\tilde{P}(k)$ at a long hold time, $t= 5$~s, when the sample is thermally equilibrated. For high $|q|$, the spectral strength in $k\leq 1/\xi_s$ is still noticeably higher than the equilibrium level of $\tilde{P}_\textrm{eq}=1$. This result is due to the presence of HQVs, which we confirmed by correlating the deviation magnitude in the low-$k$ region with the HQV number (Fig.~7).

\begin{figure}
	\centering
	\includegraphics[width=6.5cm]{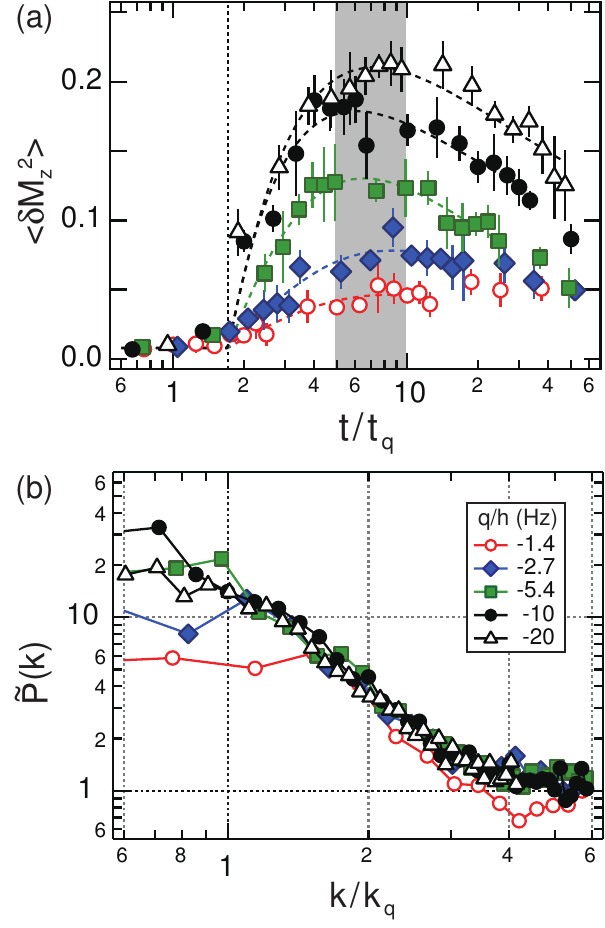}
	\caption{Time-space scaling in the spin turbulence generation. (a) Evolution of $\langle\delta M_z^2\rangle$ as a function of the rescaled time $\tilde{t}\equiv t/t_q$ with $t_q=h/\sqrt{8|q|c_2 n}$. The vertical dotted line denotes the time when $\eta=0.8$ and the grey region indicates the time window for $\langle \delta M_z^2\rangle$ to be maximized. (c) Relative spectra $\tilde{P}(k)$ with maximum $\langle \delta M_z^2\rangle$ [Fig.~4(c)] as functions of the rescaled wave number $\tilde{k}\equiv k/k_q$ with $k_q=\sqrt{2|q|m}/\hbar$.}
	\label{fig3}
\end{figure}

\subsection{Scaling behavior}

\subsubsection{Spin turbulence generation}

The spin turbulence generation is seeded by the amplification of the dynamically unstable magnon modes in the initial EAP state, in which the characteristic time and length scales are given by $t_q\sim |q|^{-1/2}$ and $l_q\sim|q|^{-1/2}$. To examine the scaling properties of the subsequent development of the spin turbulence, Fig.~5(a) displays the time evolution data of $\langle\delta M_z^2\rangle$ as functions of the rescaled time $\tilde{t}\equiv t/t_q$. For all $q$, $\langle\delta M_z^2\rangle$ begins its rapid increase at $\tilde{t}_1\approx 1.7$ and becomes maximized in the range of $5<\tilde{t}<10$. In Fig.~5(b), we display the relative power spectra $\tilde{P}(k)$ for the maximum $\langle\delta M_z^2\rangle$ as functions of the rescaled wave number $\tilde{k}\equiv k/k_q$. All the spectra collapse into a single line for $1.5<\tilde{k}<4$. This observation demonstrates that the length scale of the initial state, $l_q\sim |q|^{1/2}$, is preserved in the subsequent turbulence generation dynamics.

\begin{figure}
\centering
\includegraphics[width=7.4cm]{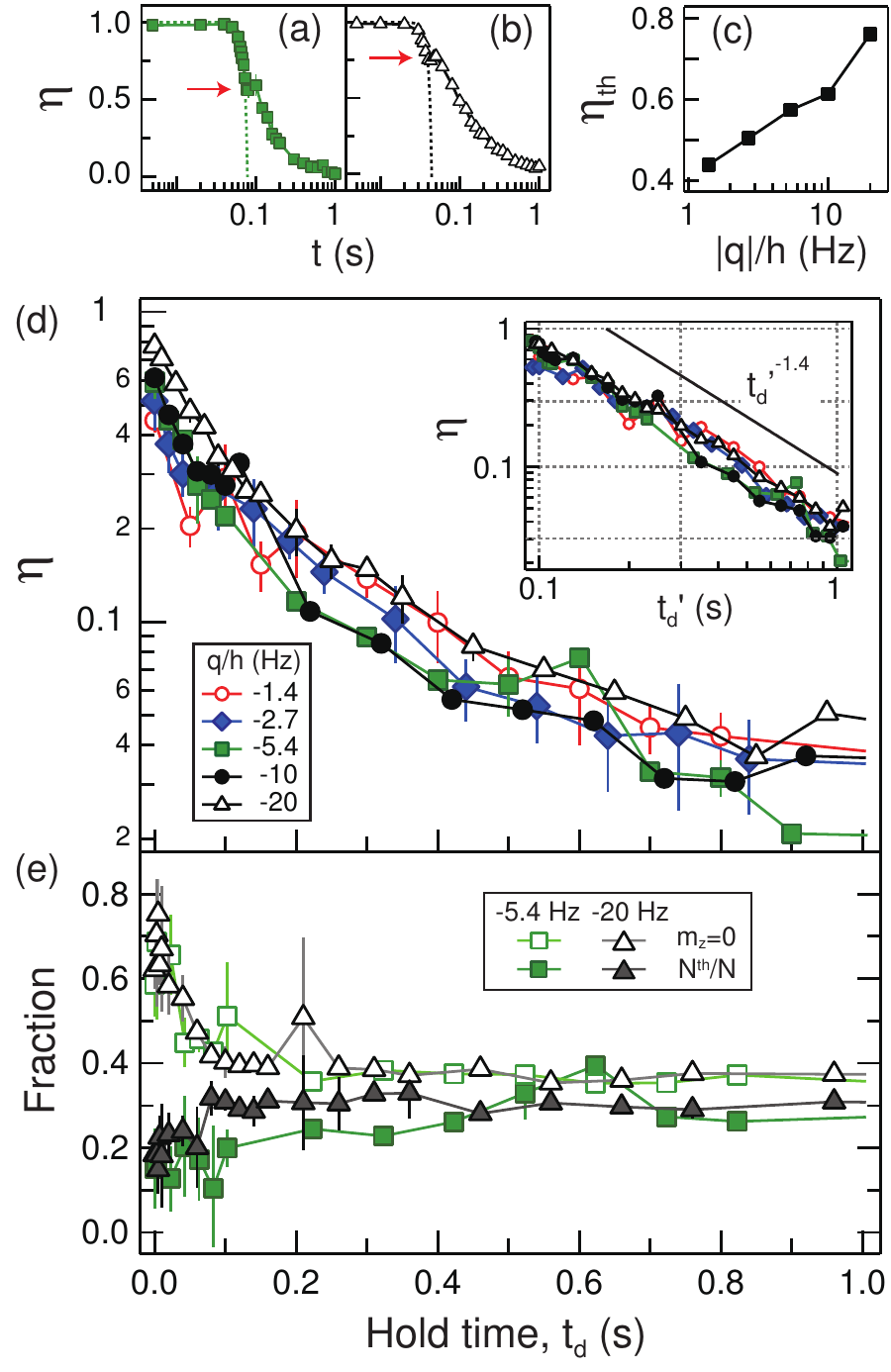}
\caption{Decay of $\eta$ in the turbulence phase. $\eta(t)$ for (a) $q/h=-5.4$~Hz and (b) $-20$~Hz, and the threshold values $\eta_{th}$ are indicated by the right arrows. (c) $\eta_{th}$ versus $|q|$. (d) Decay curves of $\eta$ as functions of $t_d=t-t_2$, where $t_2$ is the starting time of the turbulence phase. The inset shows the same data with adjusted time offsets and the solid line is a power-law fit to the data. (e) Relaxation of the thermal clouds. The solid symbols denote the thermal fraction of the sample and the open symbols denote the $m_z=0$ spin fraction of the thermal cloud. Each data point is obtained by averaging about five measurements and its error bar denotes the standard deviation of the measurements.}
\label{fig4}
\end{figure}

\subsubsection{Spin population relaxation}

As the spin turbulence develops, $\eta$ exhibits adifferent decay behavior for $\eta<\eta_{th}$ [Figs.~6(a) and 6(b)]. In Fig.~6(d), we display the decay curves of $\eta$ for various $q$ values as functions of $t_d=t-t_2$, where $t_2$ is the starting time of the turbulence phase. $\eta$ shows a nonexponential decay, where the relative decay rate $\gamma=-\frac{1}{\eta}\frac{d \eta}{dt}$ decreases as $\eta$ decreases. In the turbulence phase, the $m_z=0$ spin component, being spatially separated from the $m_z=\pm1$ components, forms axial polar domains in the condensate (Fig.~1). As the density of the $m_z=0$ atoms in the spin domains is regulated by the condensate chemical potential, a local two-body decay process would result in the constant decay of $\eta$, which cannot explain the observed nonexponential decay. The inset of Fig.~6(d) shows a log-log plot of the same data with adjusted time offsets. The determination of the exponent from a power-law fit to the data is not reliable due to its high sensitivity to the time offsets. 

In our experiments, the decay rate $\gamma$ is found to be insensitive to $q$, which is different from the general expectation of higher $\gamma$ for a higher excitation energy, $|q|$. It is speculated that the small domain size for the high $|q|$ proportionally suppresses $\gamma$, or this result may be due to the high thermal energy of the system, $k_B T \gg |q|$~\cite{Miesner99,Shin04}. In Fig.~6(e), the relaxation of the thermal cloud during the turbulence phase is characterized. The $m_z=0$ spin fraction of the thermal cloud relaxes within 0.2~s to the equilibrium level of one third, and the thermal fraction of the sample increases from $\approx 20\%$ to $\approx 30\%$. The relaxation time is compatible to the value of $1/\gamma$ at the initial decay but the decrease in $\gamma$ does not correspond with the increase of the sample temperature. The spin turbulence relaxation dynamics including the coupling to the thermal cloud~\cite{McGuirk_prl03} merits further investigation in future experiment.

\begin{figure}
\centering
\includegraphics[width=6.8cm]{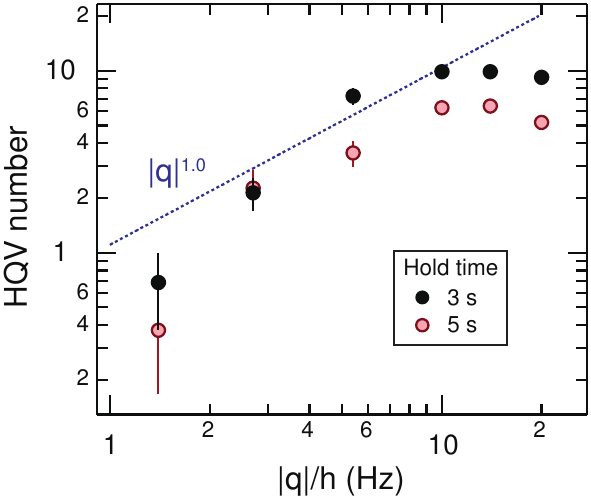}
\caption{HQV number, $N_v$, versus $|q|$. $N_v$ was measured by counting density-depleted holes in the SG images of the $m_z=\pm 1$ spin components. Each data point is the average of fifteen measurements for the same experiment and its error bars indicates the standard error of the mean. The blue dashed line denotes a power-law fit to the data in the range of $|q|/h \leq 10$~Hz for $t=3$~s.}
\label{supple2}
\end{figure}

\subsubsection{Creation of HQVs}

At long hold times of $t\geq 3$~s, when the $m_z=0$ spin component nearly vanishes in the condensate, HQVs are unambiguously identified with the density-depleted holes in the SG images (Fig.~1). We measure the HQV number, $N_v$, and find that it increases with increasing $|q|$ and saturates for $|q|/h>10$~Hz (Fig.~7). When the spatial size of the spin domains in the quenched condensate scales with $l_q$, the number of point defects created in the 2D system is expected to scale as $N_v\propto l_q^{-2}$~\cite{Bray_adv94}. A power-law fits to the data for $t=3$~s in the range of $|q|/h\leq 10$~Hz gives an exponent of $1.0\pm 0.2$, which is consistent with the length scale, $l_q\sim |q|^{1/2}$, observed in $\tilde{P}(k)$. The saturation of $N_v$ for high $|q|$ might indicate another length scale involved in the defect creation, such as the spin healing length $\xi_s$~\cite{Tomasz_prl13}, but we note that its effect was absent in our measurements of $\tilde{P}(k)$. As HQVs can be pair-annihilated in the turbulence relaxation process, leading to a non-exponential decrease of $N_v$~\cite{Kwon14}, the HQV number measured at long hold times may not linearly reflect the initial $N_v$ of the quenched BEC.

\section{Summary and outlook}

We have investigated the phase transition dynamics of quasi-2D antiferromagnetic spin-1 BECs quenched from the EAP phase to the EPP phase. We observed the emergence and decay of the spin turbulence in the quenched condensate and presented the time and space scaling properties of the phase transition dynamics near the quantum critical point. 

We can extend this work to a deeper quench regime with $|q|>2c_2 n$. In this case, the dynamic instability of the initial EAP state is driven by the magnon modes with finite wave numbers centered at $k_q=\sqrt{2m(|q|-c_2 n)}/\hbar$~\cite{Matuszewski_prl10}. It would be of great interest to examine how the energy injected at finite wave numbers flows in the subsequent evolution of spin turbulence~\cite{Fujimoto_pra16}. Because the length scale $\sim 1/k_q$ becomes comparable to and even smaller than the spin healing length $\xi_s$, it is speculated that qualitatively different turbulence states would emerge in this high $|q|$ regime. Another extension of this work is to study the effects of the quench rate across the critical point~\cite{Anquez16,Saito_pra07,Tomasz_prl13}. In particular, the power-law scaling of the HQV number after the quench is reminiscent of the Kibble-Zurek (KZ) mechanism~\cite{KZ}. We note that in the mean-field theory, the EAP-to-EPP phase transition is described as first-order; thus, the conventional KZ mechanism, which involves a continuous phase transition, cannot be applied directly to our system. A generalization of the KZ mechanism could be studied with this system. 

The turbulence of the spinor BECs represents a unique turbulence state where the mass and spin superflows are entangled. Previous turbulence studies with spinor BECs were mostly focused on the case with ferromagnetic spin interactions both experimentally~\cite{Sadler_nat06,Vengalattore_prl08,Guzman_pra11,De_pra14} and theoretically~\cite{Fujimoto_pra122,Villasenor_pra14,Williamson_prl16}. Our work demonstrates the peculiarities of the spin turbuelence in an antiferromagnetic BEC, enabling a comparative study between the ferromagnetic and antiferromagnetic cases of the spin-1 BEC system~\cite{Fujimoto_pra12,Fujimoto_pra13}.

\begin{acknowledgments}
This work was supported by Samsung Science and Technology Foundation Under Project Number SSTF-BA1601-06.
\end{acknowledgments}

\end{document}